\documentclass[12pt,epsfig,prd,showpacs]{revtex4}
\usepackage{graphicx}
\usepackage{epsfig}
\usepackage{dcolumn}
\usepackage{bm}

\newcommand{\psirww}{$J/\psi \to \gamma \omega\omega$}
\newcommand{\wppp}{$\omega \to \pi^+\pi^-\pi^0$}

\begin{document}
\vspace{-30mm}
\title{\boldmath Pseudoscalar production at $\omega\omega$
threshold in $J/\psi \to \gamma \omega\omega$}
\author{
M.~Ablikim$^{1}$,              J.~Z.~Bai$^{1}$,               Y.~Ban$^{12}$,
J.~G.~Bian$^{1}$,              X.~Cai$^{1}$,                  H.~F.~Chen$^{17}$,
H.~S.~Chen$^{1}$,              H.~X.~Chen$^{1}$,              J.~C.~Chen$^{1}$,
Jin~Chen$^{1}$
% ...
\\
\vspace{0.2cm}
(BES Collaboration)\\
\vspace{0.2cm}
{\it
$^{1}$ Institute of High Energy Physics, Beijing 100049, People's Republic of China\\
% ...
$^{12}$ Peking University, Beijing 100871, People's Republic of China\\
% ...
$^{17}$ University of Science and Technology of China, Hefei 230026, People's Republic of China\\
% ...
}
}

\vspace{0.4cm}
\date{\today}

\begin{abstract}

The decay channel $J/\psi\to\gamma\omega\omega$, $\omega\to
\pi^+\pi^-\pi^0$ is analyzed using a sample of 5.8 $\times$ 10 $^7$
$J/\psi$ events collected with the BESII detector. The $\omega\omega$
invariant mass distribution peaks at 1.76 GeV/$c^2$, just above the
$\omega\omega$ threshold. Analysis of angular correlations indicate
that the $\omega\omega$ system below 2 GeV/$c^2$ is predominantly
pseudoscalar. A partial wave analysis confirms the predominant
pseudoscalar structure, together with small $0^{++}$ and $2^{++}$
contributions, and yields a pseudoscalar mass $\rm {M}$ = 1744 $\pm$ 10 (stat)
$\pm$ 15 (syst) MeV/$c^2$, a width $\Gamma$ = $244^{+24}_{-21}$
(stat) $\pm$ 25 (syst) MeV/$c^2$, and a product branching fraction
Br($J/\psi\to\gamma\eta(1760)$) $\cdot$ Br($\eta(1760)\to
\omega\omega$) = (1.98 $\pm$ 0.08 (stat) $\pm$ 0.32 (syst)) $\times$
$10^{-3}$.
\end{abstract}
\maketitle

\section{\boldmath Introduction}
 Glueballs are expected to be copiously produced in radiative $J/\psi$
 decays. However, until now, no unique experimental signatures of such
 states have been found. The pseudoscalar ground state mesons
 ($1^1S_0$) are well established, and $\pi(1300)$, $\eta(1295)$,
 $\eta(1475)$, and $K(1460)$ are suggested as the first radial
 excitations ($2^1S_0$) of the pseudoscalar mesons~\cite{pdg}. In the
 Particle Data Group (PDG) listings, two pseudoscalar states are
 reported in the $\eta(1440)$ mass region. However, there are too many
 pseudoscalar states, and it is very difficult to find a place for 
 the lower mass $\eta(1440)$ or the $\eta(1760)$ within any $q\bar{q}$
 model~\cite{pdg}. At one time, the $\eta(1440)$ was regarded as a
 glueball candidate when it was observed in $J/\psi$ radiative
 decay~\cite{Scharre} and there was only an upper limit on its
 two-photon production~\cite{Behrend}. But this viewpoint changed when
 its radiative decay modes~\cite{radiative1,
 radiative2,radiative3,radiative4} were observed and it was also
 observed in untagged $\gamma\gamma$ collisions by the L3
 collaboration~\cite{L3}. In addition, lattice gauge theory would have
 great difficulty to accommodate such a low-mass $0^{-+}$
 glueball~\cite{lattice}.

The $\eta(1760)$ was reported by the MARK III collaboration in
$J/\psi$ radiative decays and was found to decay to
$\omega\omega$~\cite{ww-mark} and $\rho\rho$~\cite{rr-mark}.  It was
also observed by the DM2 collaboration in $J/\psi$ radiative decays in
the $\rho\rho$ decay mode with a mass of $\rm {M}$ = 1760 $\pm$ 11 MeV/$c^2$
and a width of $\Gamma$ = 60 $\pm$ 16 MeV/$c^2$~\cite{rr-dm} and in the
$\omega\omega$ decay mode~\cite{ww-dm}. The BESI experiment reported
its $\eta\pi^+\pi^-$ decay with a mass of $\rm {M}$ = 1760 $\pm$ 35
MeV/$c^2$, but without a determination of its width~\cite{bes}.  Also,
possible pseudoscalar production at threshold in the $\phi\phi$ mode
has been observed in $\pi^-p$ scattering~\cite{etkin}. The $\eta(1760)$
was suggested to be a $3^1S_0$ pseudoscalar $q\bar{q}$ meson, but some
authors suggest a mixture of glueball and $q\bar{q}$ or a
hybrid~\cite{li,wu}. Recently, in Ref.~\cite{liba}, it was argued that
the pseudoscalar glueball may be in the 1.5 to 1.9 GeV/$c^2$ mass
region, and that it also has Vector Vector decay modes.  In this paper, we
present results from an analysis of  \psirww~, \wppp~ decays, based
on a sample of 58 million $J/\psi$ events collected with the BESII
detector at the Beijing Electron-Positron Collider (BEPC). The
presence of a signal around 1.76 GeV/c$^2$ and its pseudoscalar
character are confirmed, and the mass, width, and branching fraction
are measured by partial wave analysis.

\section{ \boldmath BES detector and monte carlo simulation}
BESII is a large solid-angle magnetic spectrometer that is
described in detail in Ref.~\cite{BESII}. Charged particle momenta
are determined with a resolution of $\sigma_p/p = 1.78 \%
\sqrt{1+p^2}$ (with $p$ in GeV/$c$) in a 40-layer cylindrical
main drift chamber (MDC).  Particle identification is accomplished
using specific ionization ($dE/dx$) measurements in the MDC and
time-of-flight (TOF) measurements in a barrel-like array of 48
scintillation counters. The $dE/dx$ resolution is $\sigma_{dE/dx}$
= 8.0\%;  the TOF resolution is $\sigma_{TOF}$ = 180 ps for the
Bhabha events. Outside of the TOF counters is a
12-radiation-length barrel shower counter (BSC) comprised of gas
tubes interleaved with lead sheets. The BSC measures the energies
and directions of photons with resolutions of $\sigma_E/E
\simeq 21 \% /\sqrt{E (\rm GeV)}$, $\sigma_{\phi}$ = 7.9  mrad,
and $\sigma_z$ = 2.3 cm. The iron flux return of the magnet is
instrumented with three double layers of counters that are used to
identify muons.

In this analysis, a GEANT3 based Monte Carlo (MC) simulation program
(SIMBES) ~\cite{simbes} with detailed consideration of real
detector responses (such as dead electronic channels) is used. The
consistency between data and Monte Carlo has been carefully
checked in many high-purity physics channels, and the agreement is
quite reasonable~\cite{simbes}.

\section{\boldmath Event selection}
$J/\psi\to\gamma + 2(\pi^+\pi^-\pi^0)$ candidates are
selected from events with four charged tracks in the drift chamber and five
photons in the barrel shower counter.
\subsection{\boldmath Charged particle identification}
Each charged track, reconstructed using MDC information, is required
to be well fitted to a helix, to be within the polar angle region
$|\cos\theta| < 0.85$, to have a transverse momentum larger than 50
MeV/$c$, and have the point of closest approach of the track to the
beam axis within 2 cm of the beam axis and within 20 cm from the
center of the interaction region along the bean line.  For each track,
we make a weak particle identification requirement: either the TOF or
$dE/dx$ information must agree with that expected for a pion within four
standard deviations.

\subsection{\boldmath Photon identification}
Each candidate photon is required to have an energy deposit in the BSC
greater than 35 MeV, to be isolated from charged tracks by more than
$6^\circ$, 
to have the angle between the cluster development direction
in the BSC and photon emission direction less than  30$^\circ$,
and to have the first hit in the beginning six radiation lengths.

\subsection{\boldmath Event selection criteria}
Events are required to have four charged tracks with net charge zero
and have from 5 to 8 photon candidates. Six-constraint(6-C) kinematic fits
to the $J/\psi \to \gamma + 2(\pi^+ \pi^- \pi^0)$ hypothesis are
made with both $\gamma\gamma$ invariant masses being constrained to
the $\pi^0$ mass using all possible photon combinations.  Note that
there are 15 possible ways of combining five photons to obtain two
$\pi^0$s.  We select the combination with the highest probability and
require this probability to be greater than 10\%.  Six-constraint
kinematic fits also are applied using the $J/\psi \to 2( \pi^+
\pi^- \pi^0)$ hypothesis, and the probability of these fits is
required to be less than that of the signal hypothesis. Two further
requirements are imposed on events with more than five gammas to reduce the
background from $J/\psi \to \pi^0 2(\pi^+ \pi^- \pi^0)$.  First,
all gammas which do not belong to the chosen combination must have
$E_{\gamma} < 140$ MeV. Second, seven-constraint kinematic fits are
performed to the $J/\psi \to \pi^0 2(\pi^+ \pi^- \pi^0 )$
hypothesis with the invariant mass of the three $\gamma \gamma$ pairs
being constrained to the $\pi^0$ mass, and the event is discarded if
$P(\chi^2)_{7c} > P(\chi^2)_{6c}$.

\begin{figure}[htbp]
\centering {\includegraphics[width=4.8cm,height=5.5cm]{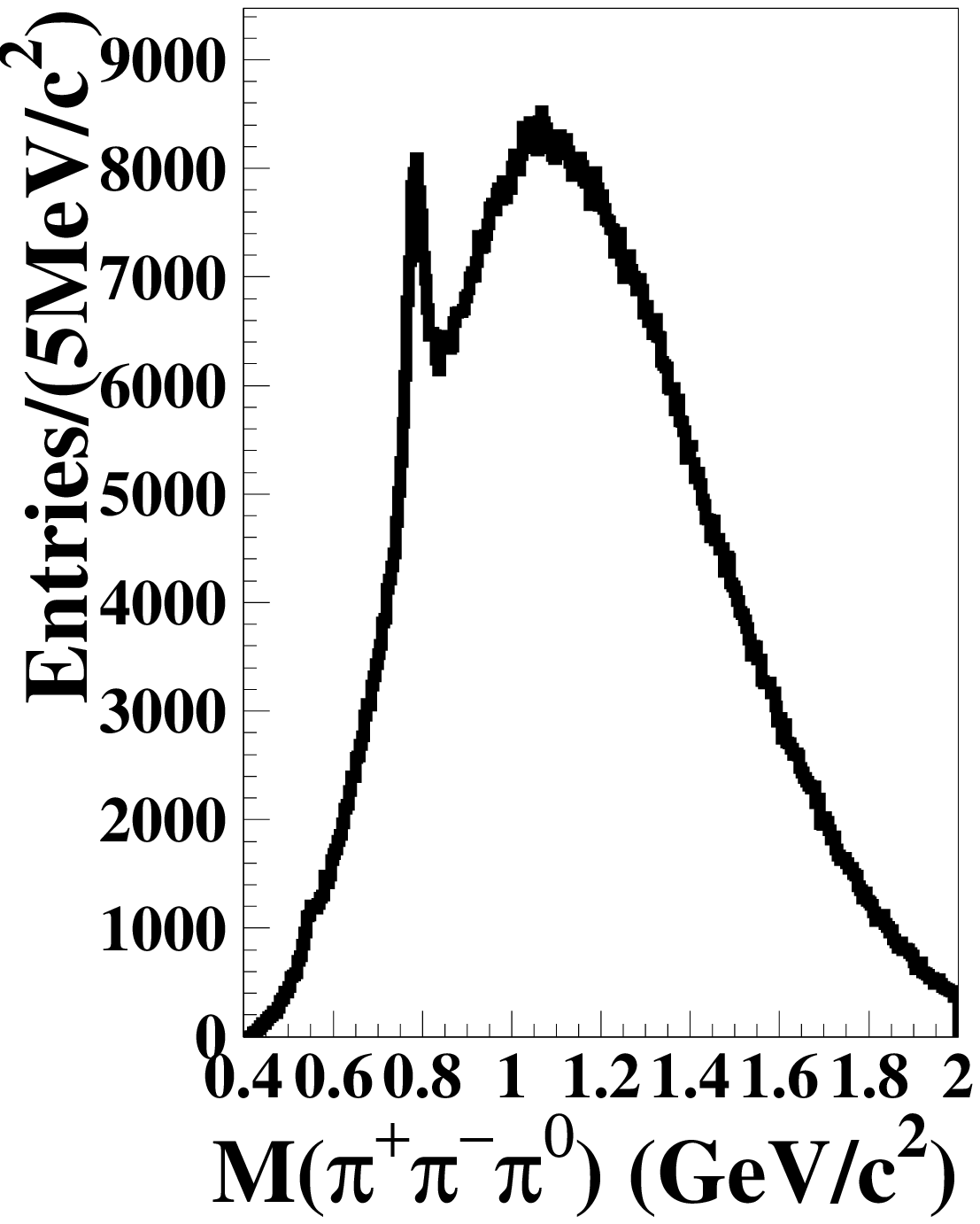}
\put(-40,120){(a)}
\includegraphics[width=4.5cm,height=5.5cm]{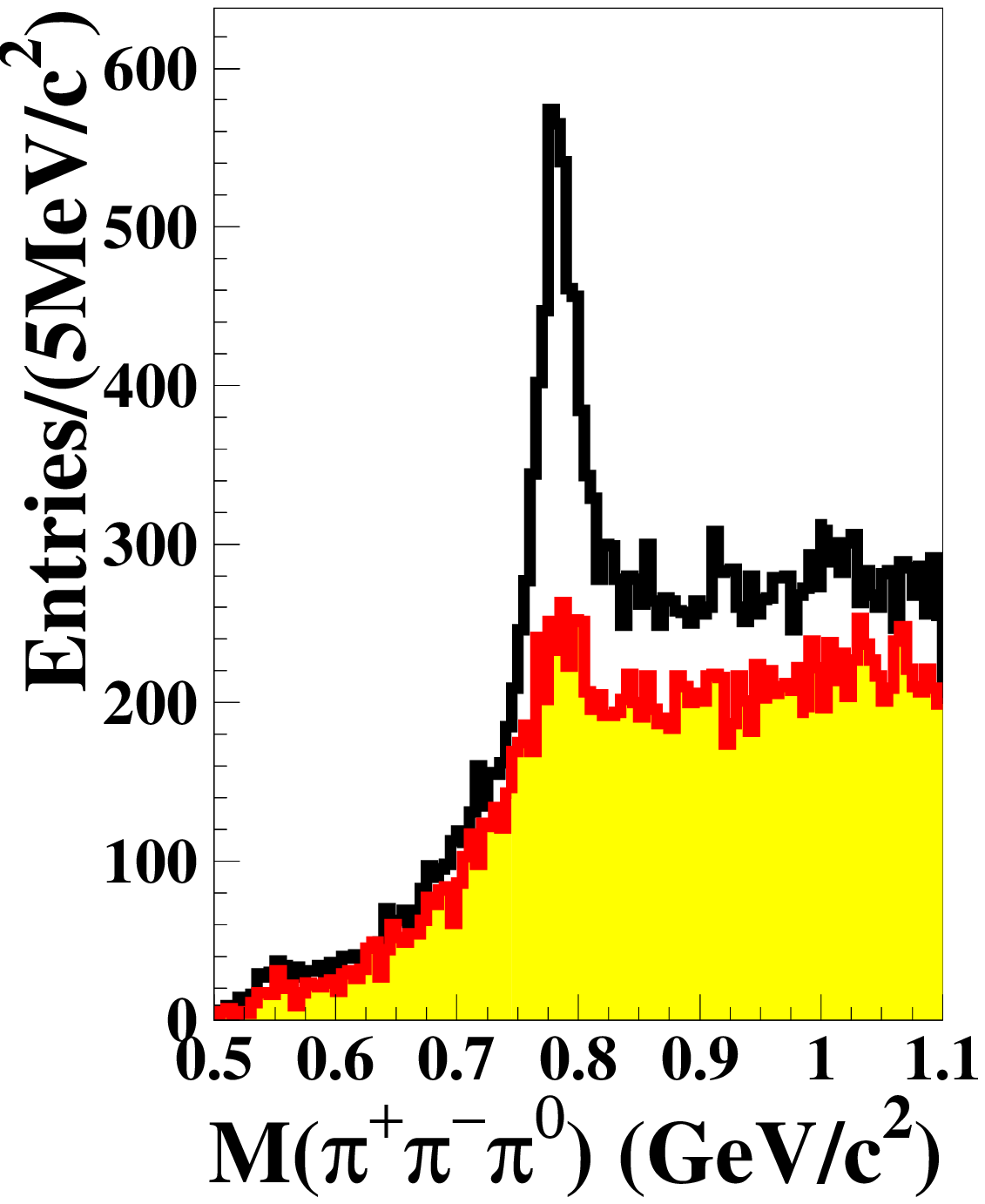}
\put(-40,120){(b)}}
\includegraphics[width=5.5cm,height=5.5cm]{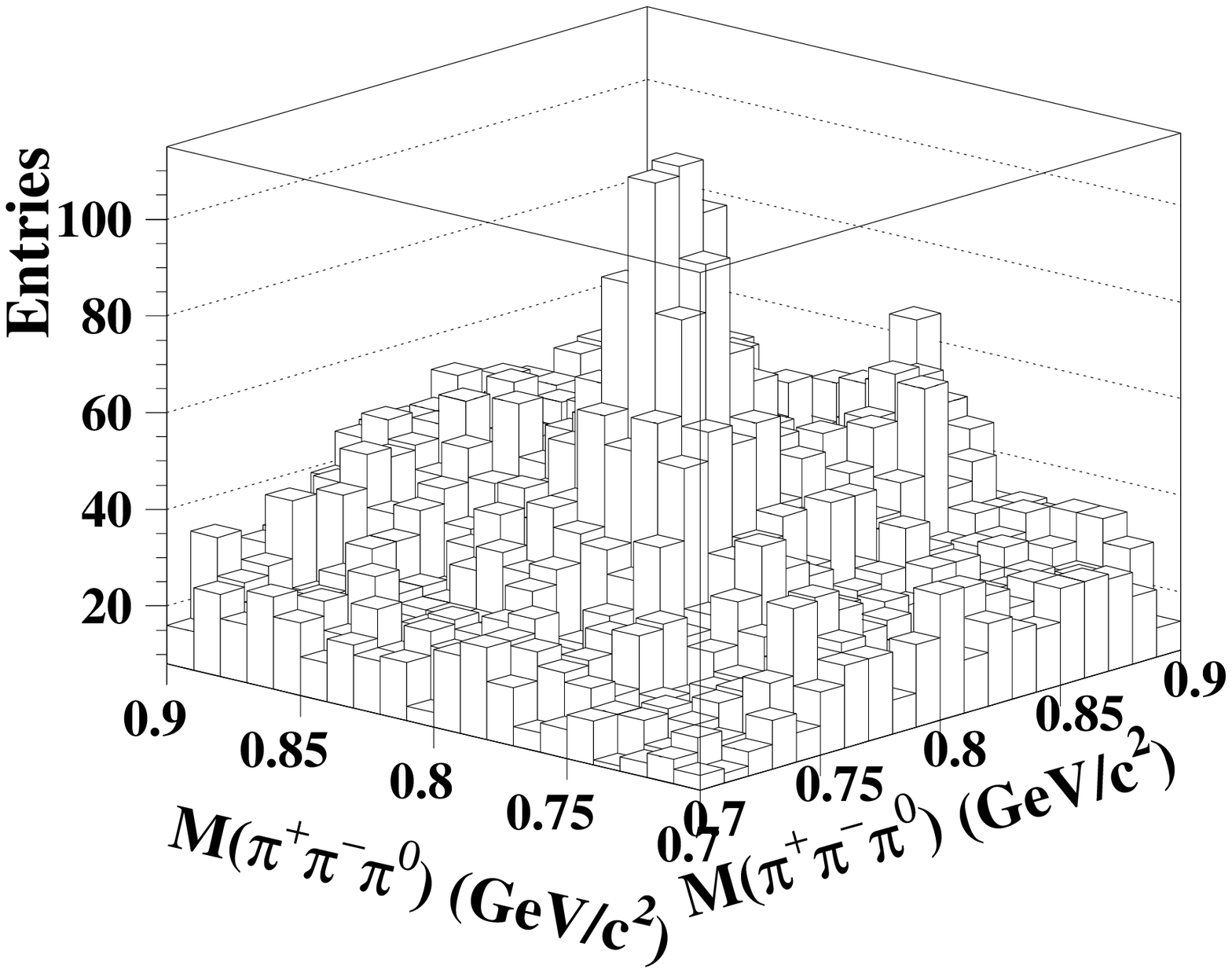}
\put(-40,120){(c)} \vskip -0.5cm \caption{(a) The $\pi^+ \pi^- \pi^0$
invariant mass distribution (8 entries per event). (b) The $\pi^+
\pi^- \pi^0$ mass distribution for the best $\omega\omega$ combination
after the requirement that the invariant mass of the other $\pi^+
\pi^- \pi^0$ is inside the $\omega$ region (open histogram), defined
by $|m_{\pi^+\pi^-\pi^0} - m_{\omega}| < $ 40 MeV/$c^2$, or in the
sideband range (shaded histogram), defined by 40 MeV/$c^2 <
|m_{\pi^+\pi^-\pi^0}-m_{\omega}|<$ 80 MeV/$c^2$. (c) The $\pi^+ \pi^-
\pi^0$ versus the $\pi^+ \pi^- \pi^0$ invariant mass (4 entries per
event).} \label{fig1}
\end{figure}

The $\pi^+ \pi^- \pi^0$ invariant mass distribution for the selected
events is shown in Fig. \ref{fig1}(a), where there are 8 entries per event. A clear
$\omega$ signal is present, mainly due to $J/\psi \to \omega \pi^+
\pi^- \pi^0 \pi^0$.  The open histogram in Fig. \ref{fig1}(b) shows
the $\pi^+
\pi^- \pi^0$ invariant mass spectrum after the requirement that the
invariant mass of the other $\pi^+ \pi^- \pi^0$ is inside the $\omega$
region ($|m_{\pi^+\pi^-\pi^0} - m_{\omega}| < $ 40 MeV/$c^2$), and the
shaded histogram is for events  after the requirement that the
invariant mass of the other $\pi^+ \pi^- \pi^0$ mass
is inside the $\omega$ sideband region (40 MeV/$c^2 <
|m_{\pi^+\pi^-\pi^0}-m_{\omega}|<$ 80 MeV/$c^2$).

\begin{figure}[htbp]
\centering
{\includegraphics[width=4.6cm,height=5.5cm]{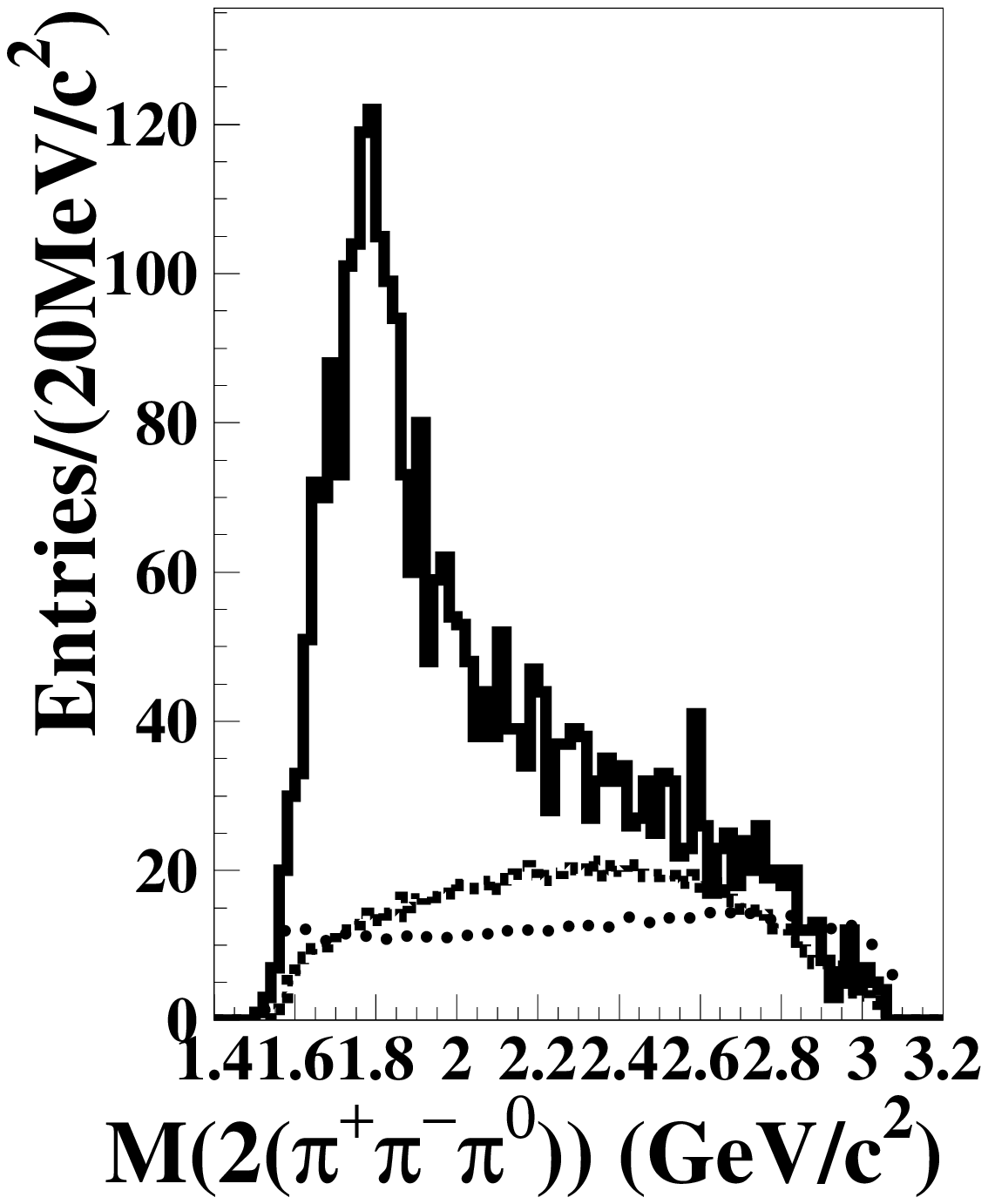}
\put(-40,110){(a)}
\includegraphics[width=5.5cm,height=5.5cm]{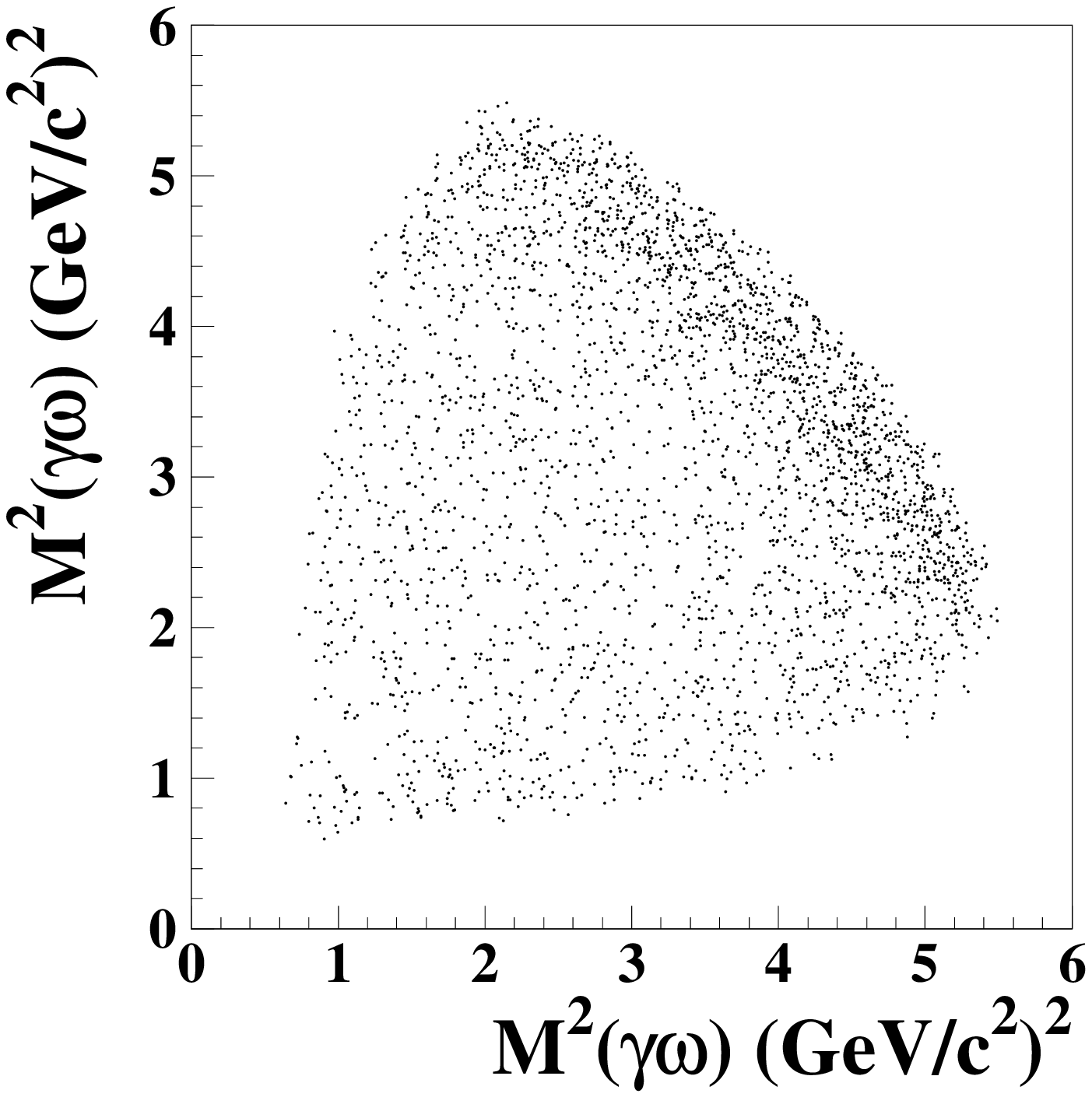}
\put(-40,110){(b)}}
\includegraphics[width=4.6cm,height=5.5cm]{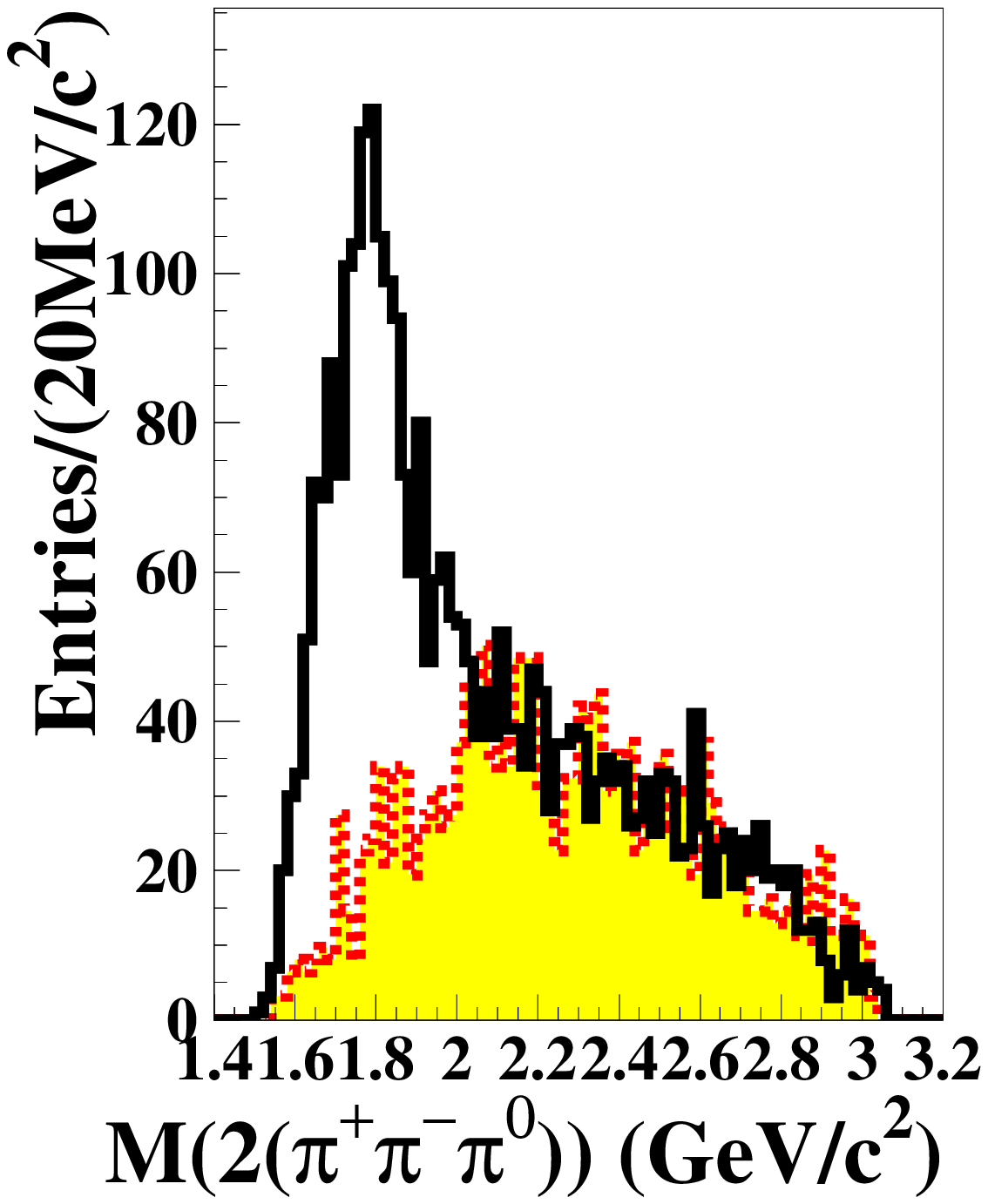}
\put(-40,110){(c)} \caption{ (a) The  $2(\pi^+ \pi^- \pi^0)$
invariant mass distribution for candidate events. The dashed
curve is the phase space invariant mass distribution, and the dotted curve
shows the acceptance versus the $\omega\omega$
invariant mass. (b) The Dalitz plot. (c) The $2(\pi^+ \pi^- \pi^0)$
invariant mass of the inclusive Monte Carlo sample (shaded histogram).} \label{fig2}
\end{figure}

The $\pi^+ \pi^- \pi^0$  versus $\pi^+ \pi^-
\pi^0$ invariant mass distribution (four entries per event) is plotted in
Fig. \ref{fig1}(c). A cluster of events is observed 
corresponding to $\omega \omega$ production.
Because the processes $J/\psi \to \omega \omega$ and
$J/\psi \to \pi^0 \omega \omega$ are forbidden by C-invariance,
the presence of two $\omega$'s is direct evidence for the
radiative decay $J/\psi \to \gamma \omega \omega$.
The histogram of Fig. \ref{fig2}(a) shows the  $2(\pi^+ \pi^- \pi^0)$
invariant mass distribution of events with both $
\pi^+ \pi^- \pi^0$ masses within the $\omega$ range ($|m_{
\pi^+\pi^-\pi^0 }-m_{\omega}| <$40 MeV/$c^2$). There are 3046
events with a
clear peak at 1.76 GeV/$c^2$. The phase space invariant mass
distribution 
and the acceptance versus $\omega \omega$ invariant mass
are also shown in the figure. The corresponding Dalitz plot is
shown in Fig. \ref{fig2}(b).

From MC simulation, the $\omega$ signal can be well fitted with a
double Gaussian with widths 14.6 MeV/$c^2$ and 45.5 MeV/$c^2$, which
makes  it
difficult to evaluate the background from the sideband events
especially for the events near $\omega\omega$ threshold. To ensure
that the structure at the $\omega\omega$ mass threshold is not due to
background, we have made studies of potential background
processes. The main background sources come from
$J/\psi \to\omega \pi^+ \pi^- \pi^0 \pi^0 $, $\gamma 2(\pi^+ \pi^- \pi^0)$,
and $\pi^0 2(\pi^+ \pi^- \pi^0)$. More than one half of the
background comes from the first case. However, none of these background
channels gives a peak at 1.76 GeV/$c^2$ in the $2(\pi^+ \pi^- \pi^0)$
 invariant mass spectrum. In addition,
possible backgrounds were checked with a MC sample of 58 million
inclusive $J/\psi$ decays generated by the LUND model~\cite{chenjc},
with $\gamma \omega \omega$ events removed. The shaded histogram of
Fig. \ref{fig2}(c) shows the $2(\pi^+ \pi^- \pi^0)$ invariant mass distribution of the inclusive sample. There
is no peak at the $\omega\omega$ mass
threshold in the invariant mass  distribution at around 1.76
GeV/$c^2$, while the inclusive MC distribution is
comparable with data for  $2(\pi^+ \pi^- \pi^0)$ invariant
mass greater than 2 GeV/$c^2$. A background evaluation is performed by
fitting the $\pi^+ \pi^- \pi^0$ mass distribution for events with the
other   $\pi^+ \pi^- \pi^0$
within the $\omega$ signal region (Fig. \ref{fig1}(b)).  The background shape is
obtained from the inclusive MC sample, and the signal shape is obtained from the phase
space MC sample of $J/\psi \to \gamma \omega \omega$.  The number of events is
free in the fitting.  The fitting
yields 1441 $\pm$ 50 background events within the
$\omega \omega$ invariant mass range from 1.6 GeV/$c^2$ to 2.8 GeV/$c^2$.

\section{\boldmath Angular correlation analysis}
An analysis of the angular distributions of the accepted events
has been performed in order to estimate whether the $\omega\omega$
production below 2 GeV/$c^2$ belongs to a resonant state with
definite spin-parity. Candidate events and side-band background events
are analyzed choosing the $\omega\omega$ pair whose quadratic sum of the two
differences $(m_{\pi^+\pi^-\pi^0}-m_{\omega})$ is minimum. For
systems of two vector mesons, the distribution of $\chi$, the
azimuthal angle between the normals to the two $\omega$ decay
planes in the $\omega \omega$ rest frame, provides a unique
signature for the spin and parity~\cite{chi1,chi2,chi3,chi44}. The
distribution takes the form $dN/d{\chi} \propto 1 + \beta
\cos(2\chi)$, where $\beta$ is a constant which is independent of
the polarization of the $\omega \omega$ system, but exhibits
strong correlation with the spin-parity. $\beta$ is zero for odd
spin and non-zero for even spin. Its sign is the parity of the
$\omega \omega$ system. For $J^P = 0^-$, where $\beta$ is $-1$,
$dN/d{\chi} \propto \sin^2{\chi}$, and the effect is maximal.
\begin{figure}[htbp]
\centering
{\includegraphics[width=6.0cm,height=6.5cm]{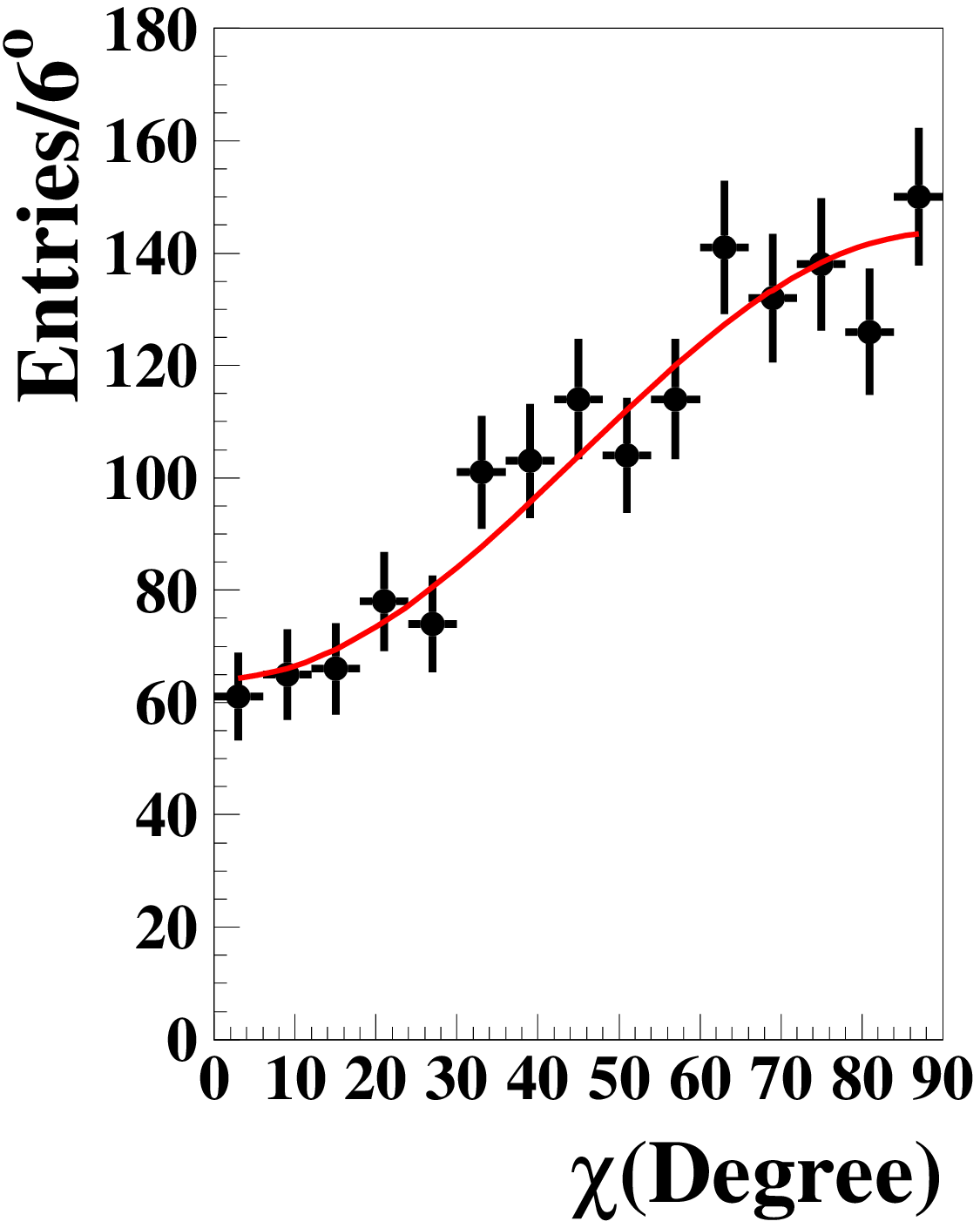}
\put(-120,150){(a)}
\includegraphics[width=6.0cm,height=6.5cm]{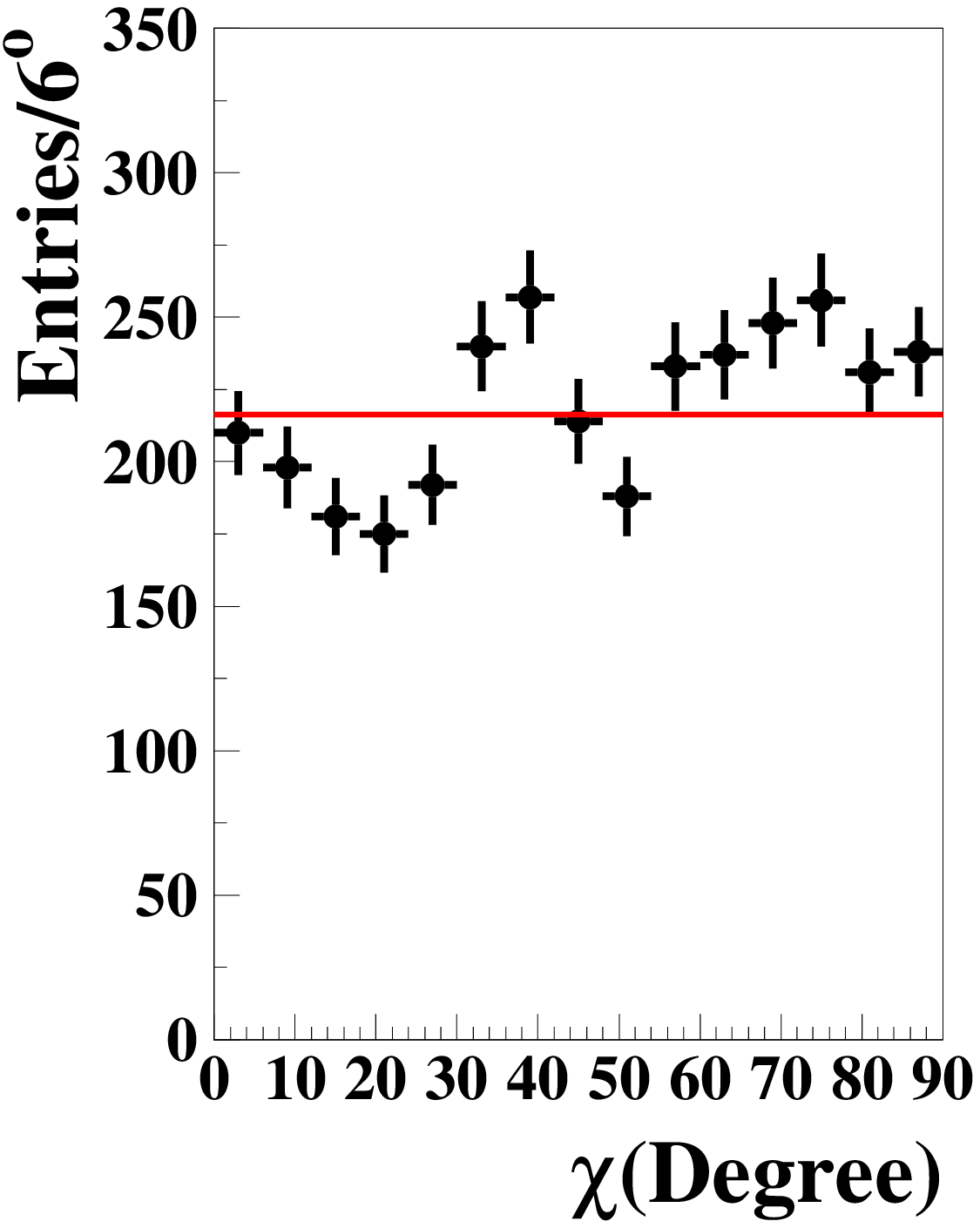}
\put(-120,150){(b)}}
\caption{ Distribution of $\chi$, the azimuthal angle between the
normals to the  two $\omega$ decay planes for
(a) for $\omega-\omega$ signal events ($m_{\omega\omega} < 2$ GeV/$c^2$)
 and (b) for $\omega$ sideband events ($m_{\omega\omega} < 2$ GeV/$c^2$)}
\label{chi}
\end{figure}

Fig. \ref{chi} shows the $\chi$ distribution for a) $J/\psi \to
\gamma \omega \omega$ events with $m_{\omega\omega}$ less than 2
GeV/$c^2$, and b) events from the $\omega$ sidebands, where
$\omega$ sideband is defined with a least one $m_{ \pi^+ \pi^-
\pi^0}$ mass in the range of 40 MeV/$c^2<|m_{\pi^+ \pi^- \pi^0} -
m_{\omega}|<120$ MeV/c$^2$. The distributions for signal and
background events are strikingly different, indicating a large
component with even spin and odd parity in the signal region. The
solid line in Fig. \ref{chi}(a) is the result of a fit to $a + b
\sin^2\chi$, which yields a $\sin^2 \chi$ contribution of the
$\gamma \omega\omega$ event candidates below 2 GeV/$c^2$ of
$38.3\pm3.5 \%$.

\section{\boldmath Partial wave analysis}
A partial wave analysis (PWA) has been carried out for events with
$2(\pi^+\pi^-\pi^0)$ invariant mass from 1.6 GeV/$c^2$ to 2.8
GeV/$c^2$. The sequential decay process can be described by $J/\psi
\to \gamma X, X\to \omega\omega$ and $\omega \to
\pi^+\pi^-\pi^0$. The amplitudes of the two body or three body decays
are constructed using the covariant helicity coupling amplitude
method~\cite{wuning1}. The intermediate resonance $X$ is denoted with
the normal Breit-Wigner propagator $\rm { BW } = 1/(\rm{ M }^2-s-i \rm{ M }\Gamma)$, where
$s$ is the $\omega\omega$ invariant mass-squared  and $\rm M$ and
$\Gamma$ are the resonance's mass and width. The amplitude of
sequential decay process is the product of all decay amplitudes and
the Breit-Wigner propagator.

The $\chi$ angular distribution  shows a strong contribution from
structures with  even spin and odd parity
for $\omega\omega$ invariant mass below 2 GeV/$c^2$. Therefore
the study of the $\eta(1760)$ is the main goal of this analysis. From
the PDG, three $f_2$ resonances can decay into $\omega\omega$
final states, $f_2(1560)$, $f_2(1640)$, and $f_2(1910)$. Because the
mass of $f_2(1560)$ and $f_2(1640)$ are very close,  only one
resonance, $f_2(1640)$, is considered in the analysis. The $f_0(1710)$
is a well known resonance, and its spin-parity allows it to
decay to a $\omega\omega$ final state, so it is included in the
fit. Finally, four possible intermediate  resonances
$\eta(1760)$, $f_0(1710)$, $f_2(1640)$, and $f_2(1910)$ are included in
the final analysis,  and the total differential cross section
$d\sigma/d\Phi$ is 

\begin{eqnarray}
\frac{d\sigma}{d\Phi} =| A(\eta) + A(f_0) + A(f_2^1) + A(f_2^2)|^2 + BG,
\end{eqnarray}
where $A(\eta)$ , $A(f_0)$,  $A(f_2^1)$, and $A(f_2^2)$ are the
total amplitudes of the resonances $\eta(1760)$, $f_0(1710)$,
$f_2(1640)$, and $f_2(1910)$, respectively, and $BG$ denotes the
background contribution, which is described by phase space. 

The relative magnitudes and phases of the amplitudes are determined by
an unbinned maximum likelihood fit. 
The basis of likelihood fitting is calculating the probability that 
a hypothesized probability distribution function would produce the 
data set under consideration. The joint probability density for 
observing the N events in the data sample is 
\begin{eqnarray}
  \mathcal{L} = \prod_{i=1}^{N}{P(x_i)}
\end{eqnarray}
where $P(x_i)$ is the probability to produce event $i$ characterized 
by the measurement $x_i$, which is the normalized differential cross
section:
\begin{eqnarray}
  P(x_i) = \frac{\left( \frac{d\sigma}{d\Phi} \right)_i}
   {\int\frac{d\sigma}{d\Phi}{d\Phi}}
\end{eqnarray}
The normalization integral $\int\frac{d\sigma}{d\Phi}{d\Phi}$ is 
done by the phase space MC sample, the details are described in
Ref.~\cite{guozj}. The free parameters are optimized by MINUIT
~\cite{minuit}. Technically, rather than maximizing $\mathcal{L}$,
the $\mathcal{S} = -\rm{ln} \mathcal{L}$ is minimized. In the 
minimization procedure, a change in log likelihood of 0.5 represents 
a one standard deviation effect for one parameter case.

\begin{figure}[htbp]
\begin{center}
{\includegraphics[width=15.5cm,height=11.5cm]{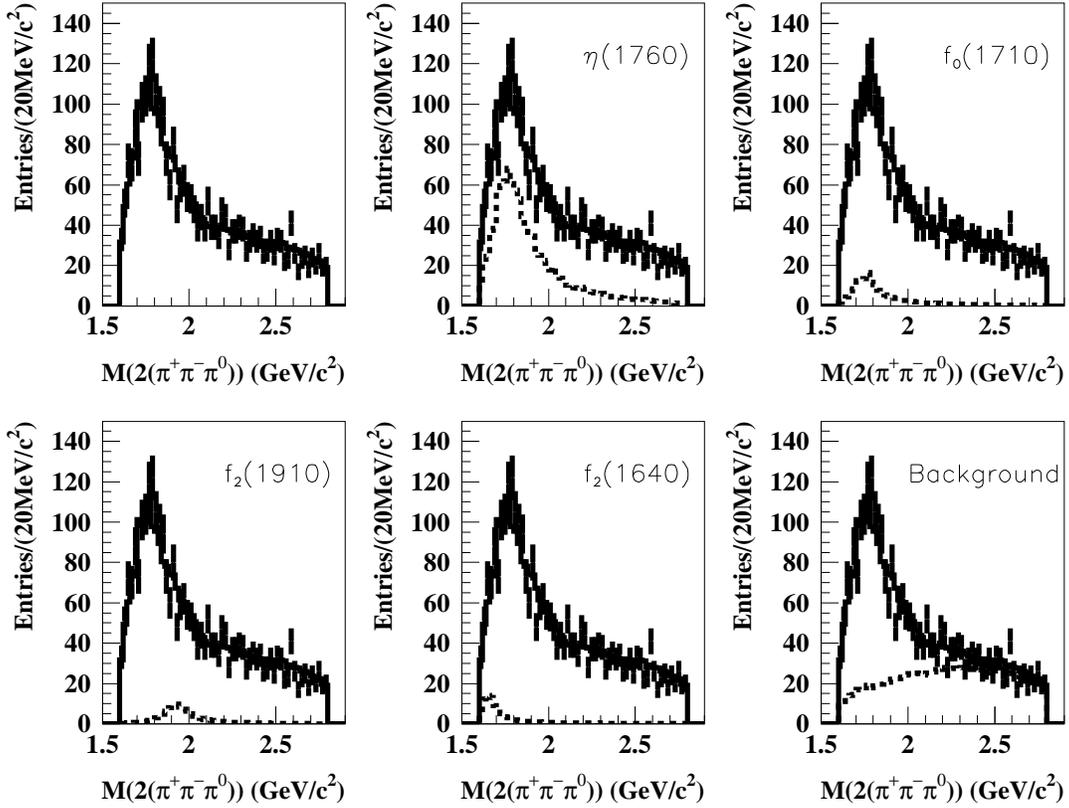}}
\caption{The $2(\pi^+\pi^-\pi^0)$ invariant mass distribution for
$J/\psi\to \gamma \omega\omega$. The points with error bars are data,
the full histograms show the projection of the maximum likelihood fit, and
the dashed histograms show the contributions of each of the resonances
and background. }
\label{mass}
\end{center}
\end{figure}

For the production of a pseudoscalar, only $\mathcal{P}$
waves are allowed in both the radiative decay $J/\psi \to \gamma X$
and the hadronic
decay $X\to \omega\omega$. For the production of a scalar, both
$\mathcal{S}$ and $\mathcal{D}$ waves are possible in both the
radiative and hadronic decays, but only $\mathcal{S}$ wave
is considered in the fit. For the production of a $2^+$ resonance,
$\mathcal{S}$ waves in both decays are considered, and two  of
three $\mathcal{D}$ waves in the radiative decay and only one
$\mathcal{D}$ wave in the hadronic decay, corresponding to the
lower overall spin of the $\omega\omega$ system, are considered.
From the analysis of angular correlations, it is found that the
contributions from $f_0(1710)$, $f_2(1640)$, and $f_2(1910)$ are very
small, so the mass and width of these resonances are fixed to PDG
values, but the amplitudes are allowed to vary in the fit. The
mass and width of the $\eta(1760)$ are obtained from the optimization; the
mass and width are $\rm {M}$ = 1744 $\pm$ 10 MeV/$c^2$ and $\Gamma$ =
$244^{+24}_{-21}$ MeV/$c^2$, where the errors are statistical. The final global
fit, and the contributions of all resonances and backgrounds are
shown in
Fig. \ref{mass}.

\begin{figure}[htbp]
\centering {\includegraphics[width=15.5cm,height=11.5cm]{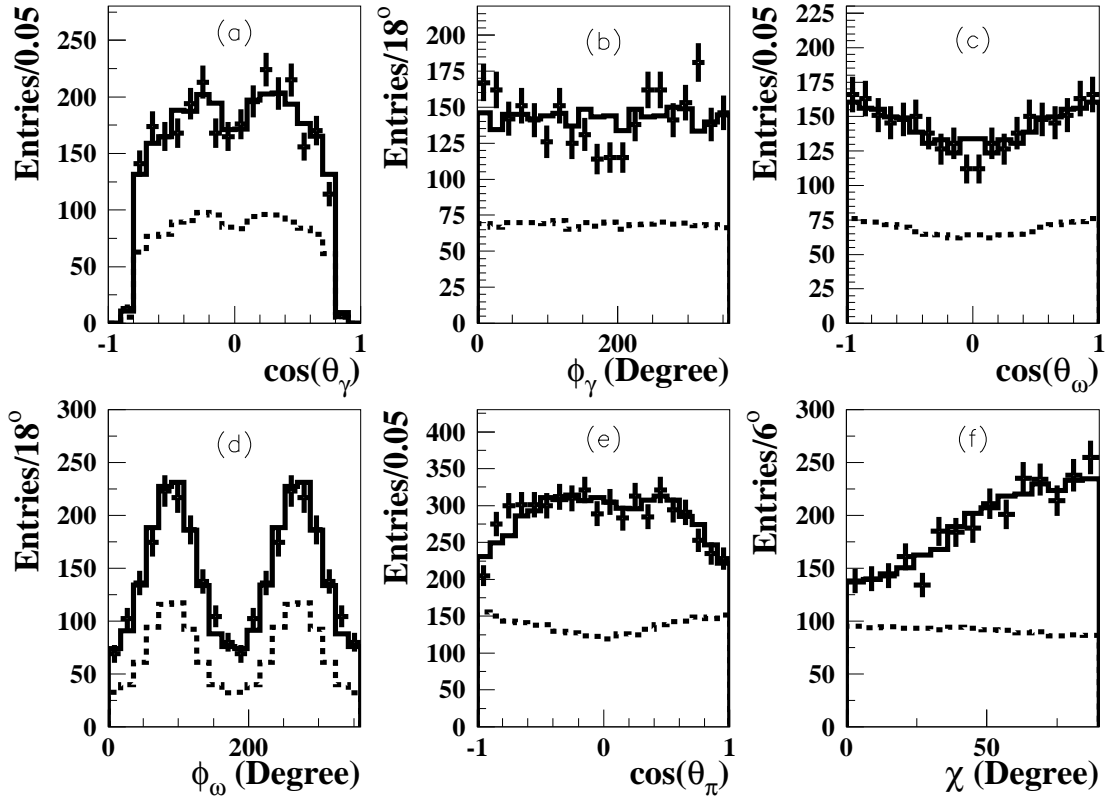}}
\caption{Comparisons of angular distributions between data and fit
projections of the global fit. The dashed histograms show the background
contributions. (a, b) The polar and azimuthal
angles of the radiative gamma. (c, d) The polar and azimuthal angles of the
$\omega$ in the $\omega\omega$ rest system. (e) The polar angle of
the normal to $\omega$ decay plane in the $\omega$  rest system,
both $\omega$'s angle are projected in the same plot. 
(f) The $\chi$ distribution.
}
\label{ang}
\end{figure}

Comparisons of angles of fit projections and data
are shown in Fig. \ref{ang}.
To determine the goodness of fit, a $\chi^2$ is calculated
by comparing the data and fit projection histograms,
where $\chi^2$ is defined as~\cite{guozj}:
\begin{eqnarray}
\chi^2 = \sum_{i=1}^{N}\frac{(n_i - v_i)^2}{v_i}
\end{eqnarray}
and $n_i$ and
$v_i$ are the number of events for data and the fit projections in the
$i$th bin of each figure, respectively.  The $\chi^2$ and number
of degrees of freedom ($ndf$) for the $\omega\omega$ invariant mass and the
angle distributions are shown in Table~~\ref{table1}, where the number
of bins is taken as the number of degrees of freedom. The values of
$\chi^2/ndf$ are between
0.6 and 1.8, indicating good agreement between data and the fit.
\begin{table}[htbp]
\begin{center}
\begin{tabular}{lccc  c  c  c  c} \hline\hline
Variable  &~~mass~~&~~~$\theta_\gamma$~~~&~~~$\phi_\gamma$~~~ &
~~~$\theta_\omega$~~~&~~~$\phi_\omega$~~~&~~~$\theta_\pi$~~~&~~~$\chi$~~~ \\ \hline
$\chi^2$  &  ~68.2 & ~20.3& ~35.2& ~12.9&14.7 &~22.5& ~13.4\\ \hline
$ndf$       &  ~~~60 & ~~~18& ~~~20& ~~~20&~~20 &~~~20& ~~~15\\ \hline
$\chi^2/ndf$   & ~1.14  & ~1.13&~1.76 & ~0.65&~0.74&~1.13&~0.89\\ \hline
C.L. (\%)&  ~21.9 & ~31.6& ~~1.9& ~88.2&~79.3& ~31.4& ~57.1\\ \hline
\end{tabular}
\caption{Goodness of fit check for the invariant mass distribution and
  angular distributions shown in Fig.~\ref{ang}, where $ndf$ and C.L.
are the number of degrees of freedom and the corresponding confidence level.}
\label{table1}
\end{center}
\end{table}

The numbers of events, detection efficiencies, and the corresponding
branching fractions for $J/\psi \to \gamma X \to \gamma \omega\omega$
with intermediate resonances $\eta(1760)$, $f_0(1710)$, $f_2(1640)$,
and $f_2(1910)$ are shown in the Table \ref{table2}, where the errors
are statistical errors only and the correlations between the different
resonances are included.  The magnitudes and phases of the
partial amplitudes from the PWA are used in the detection efficiency
determination. Details of the fitting procedure and the detection
efficiency determination can be found in Ref.~\cite{guozj}. The
changes of the log likelihood value $\Delta \mathcal{S}$ when the
corresponding resonance is dropped from the fit and the statistical
significance for each component are also shown in 
Table~\ref{table2}, where the significance is calculated from
the difference between  $\mathcal{S}$ values of the fits
with and without the resonance. The product
branching fraction is Br($J/\psi\to\gamma\eta(1760)$) $\cdot$
Br($\eta(1760)\to \omega\omega$) = (1.98 $\pm$ 0.08 (stat))
$\times$ 10$^{-3}$, and the statistical significance of the
$\eta(1760)$ is above 10 $\sigma$. All the resonances listed improve
the fitting by more than 5$\sigma$. If the spin-parity of $\eta(1760)$
is replaced by $0^{++}$ in the fit, the log likelihood is worse by
248.0, so the possibility that its spin-parity is $0^{++}$ is excluded
by at least 10 $\sigma$.
\begin{table}[htbp]
\begin{center}
\begin{tabular}{c|cccccc} \hline\hline
~resonance~~&~~Events~~&~eff( \% )~&~~$Br$($ \times 10^{-3}$)~~&~~Sys Err( \% )
&~~$\Delta\mathcal{S}$~~&Sig. \\ \hline
$\eta(1760)$&~$1045\pm41$~&~1.15~&~~$1.98\pm0.08$&~16.4~&280 & $>10\sigma$\\
$f_0(1710)$ &  $180\pm37$ &~1.27~&~~$0.31\pm0.06$&~25.1~&~23.5 & ~$6.5\sigma$\\
$f_2(1910)$ & $151\pm32$  &~1.68~&~~$0.20\pm0.04$&~64.9~&~23.5 & ~$5.8\sigma$\\
$f_2(1640)$ & $141\pm26$  &~1.08~&~~$0.28\pm0.05$&~59.6~&~21.4 & ~$5.5\sigma$\\
\hline
\end{tabular}
\caption{The fitted number of events, detection efficiency, product branching
fraction,  systematic error, log likelihood value differences, and statistical significance
of each resonance.}
\label{table2}
\end{center}
\end{table}

The fit determines 1371 $\pm$ 45 background events, which is
consistent with the result obtained from Fig. \ref{fig1}(b).  Another
technique for treating background, which was used in
Refs.~\cite{guozj, dongly, wu1}, is to set $BG$ to 0 in Eq. (1)
and to cancel the background contribution by including MC data in the
fit with the opposite sign of log likelihood compared to the data. As a check we have also used this method. The
MC sample is obtained using the inclusive MC, and the number of
background events is fixed to the fitting results of
Fig. \ref{fig1}(b). The
$\eta(1760)$ mass and width obtained are 1742 $\pm$ 10 MeV/$c^2$ and
234 $\pm$ 17 MeV/$c^2$, respectively, and the product branching fraction of
$J/\psi \to \gamma X, X\to \omega\omega$ is (2.01 $\pm$ 0.08) $\times$
$10^{-3}$.

Other states listed in the PDG between 1.6 GeV/$c^2$ and  2.0
 GeV/$c^2$, that are consistent with decay into $\omega \omega$ under spin-parity
 constraints, are the $\eta_2(1645)$, $\eta(1870)$, $f_2(1810)$, $f_2(1950)$.
 If these resonances are included in the fit, the log
 likelihood values $\mathcal{S}$ improve by 6.2, 9.0, 8.9, and 8.4,
 respectively, while the $\eta(1760)$ masses, widths, and branching
 fractions are consistent with the final fit result within statistical
 errors.  The difference between results including and not including
 the $\eta(1870)$ will be taken as a systematic error.

The $f_0(1790)$ has been recently claimed in $J/\psi$
 decay~\cite{dongly}. If the parameters of $f_0(1710)$ are replaced
 with those of the $f_0(1790)$, the log likelihood value is improved
 by 2.8 after the reoptimization. 
 The reoptimized mass, width, and product branching ratio of 
$\eta(1760)$ are $1744 \pm 10$ MeV/c$^2$, $238 \pm 20$ MeV/c$^2$,
and $(1.97 \pm 0.07) \times 10^{-3}$, respectively, and the product 
branching ratio of $f_0$ is then $(0.39 \pm 0.07) \times 10^{-3}$ 
(statistical error only).
 Recently a scalar enhancement, the $f_0(1812)$, near $\omega\phi$
 threshold in $J/\psi\to\gamma\omega\phi$ decay was reported by the
 BESII collaboration~\cite{penghp}, and information on the
 corresponding $\omega\omega$ decay mode is very important to
 understand its nature~\cite{zhao,ppp,Liba2,bug,hexg1,hexg}. If the $f_0(1710)$
 parameters are replaced with the $f_0(1812)$~\cite{penghp}, the log
 likelihood value is improved by 0.8 after 
the reoptimization.
 The reoptimized mass, width, and product branching ratio of
$\eta(1760)$ are $1740 \pm 10$ MeV/c$^2$, $246 \pm 24$ MeV/c$^2$,
and $(1.97 \pm 0.07) \times 10^{-3}$, respectively, and the product 
branching ratio of $f_0$ is then $(0.26 \pm 0.05) \times 10^{-3}$ 
(statistical error only).
If both the $f_0(1710)$ and
 $f_0(1812)$ are added in the fitting, the log likelihood values is
 improved by 5.2. If the $f_0(1710)$'s parameters are replaced with
 those of the $f_0(2020)$, the log likelihood value will be improved
 by 2.9. But if no scalar in this energy region is used in the fit,
 the log likelihood value is worse by 23.5, corresponding to 6.5
 $\sigma$. From these tests, we conclude that a scalar is needed,
 but it is very difficult to determine its mass and width accurately
 due to the dominant contribution of the pseudoscalar. If the
 parameters of the $f_2(1640)$ are replaced with those of $f_2(1560)$,
 the likelihood value is improved by 0.55 after mass and width
 reoptimization. If the parameters of the $f_2(1910)$ are replaced by
 those of the $f_2(1950)$, the log likelihood value is improved by
 2.1. In the final fit, the $\omega$ decay amplitude is described with
 sequential two body decays with $\mathcal{P}$ wave. If the $\omega$
 decay amplitude is taken to be constant, the results do not change
 much. In all these tests, the $\eta(1760)$ masses, widths, and the
 branching fractions are consistent with the final fit results. The
 differences are included in the systematic errors.

\section{ \boldmath Systematic error}
The systematic errors are estimated by considering the following: (a)
The $f_0(1710)$ is replaced with the $f_0(2020)$. (b) The $f_2(1640)$
is replaced with the $f_2(1560)$. (c) The $f_2(1910)$ is replaced with
the $f_2(1950)$. (d) The fit is done with and without the
$\eta_2(1870)$. (e) A constant $\omega$ amplitude is used in the
fit. (f) Different background treatments. (g) Different $\gamma$
selection criteria: energy greater than 50 MeV/$c^2$, and the minimum
angle between the gamma and the nearest charged track greater than
$10^\circ$. (h) Changing the polar angle requirement of charged tracks
to $|\cos\theta| < 0.8$. (i) Changing the 6-C kinematic fit
probability requirement from Prob$_{6c}>0.1$ to Prob$_{6c}>0.05$. (j)
Changing the $\pi^+\pi^-\pi^0$ invariant mass requirement from the 40
MeV/$c^2$ to 45 MeV/$c^2$. The total errors are obtained by adding the
individual errors in quadrature. The total mass and width systematic
errors are 0.83\% and 10.5\%, respectively. For the branching fraction
systematic error, the uncertainties in the MDC tracking, the photon
identification
efficiency, the $\omega\to \pi^+\pi^-\pi^0$ branching fraction, and
the number of $J/\psi$ events are also included, and the total
branching fraction systematic errors are 16.4\%, 25.1\%, 64.9\%, and
59.6\% for $\eta(1760)$, $f_0(1710)$, $f_2(1910)$, and $f_2(1640)$,
respectively, which are also listed in Table \ref{table2}.

\section{ \boldmath Discussion}
The $\eta(1760)$ is prominently produced in $J/\psi \to \gamma \omega
\omega$.  Its two-gluon coupling can discriminate between its gluonic
and $q\bar{q}$ nature~\cite{li,wu}. If
perturbative QCD works well and the non-relativistic approximation is
applicable, the formalism proposed in Refs.~\cite{close,cakir}, which
connects the two-gluon width $\Gamma(\eta(1760) \to gg)$ of
$\eta(1760)$ to the radiative $J/\psi$ branching fraction, $Br$($J/\psi
\to \gamma\eta(1760)$), can be used (see Ref.~\cite{close}, Eq. 3.4):
\begin{eqnarray}
10^3 Br(J/\psi \to \gamma \eta(1760)) = \left(\frac{\rm M}{1.5 {\rm
   ~ GeV}/c^2}\right)\left(\frac{\Gamma_{\eta(1760)\to gg}}{50 {\rm
      ~MeV}/c^2}\right) \frac{x |H_{ps}(x)|^2}{45},
\end{eqnarray}
where $\rm {M}$ is the $\eta(1760)$ mass, $\Gamma_{\eta(1760)\to gg}$ is
the width to $gg$, and $x |H_{ps}(x)|^2$ is the magnitude of the loop
integral calculated in Ref.~\cite{close}.  Therefore, the gluonic
content of $\eta(1760)$ is estimated by its two-gluon coupling, which
is calculated from its mass, width, and branching fraction in
$J/\psi$ radiative decay.  Rewriting $\Gamma_{\eta(1760)\to gg}$ as
$\Gamma_{\eta(1760)} \cdot Br({\eta(1760)\to gg)}$ and multiplying both
sides of the equation by $Br(\eta(1760)\to \omega\omega)$, we obtain:
\begin{eqnarray*}
\lefteqn{Br(\eta(1760)\to gg) \cdot Br(\eta(1760)\to\omega\omega)} \\
&=& 10^3\left[ Br(J/\psi \to \gamma \eta(1760))\cdot Br(\eta(1760)\to
\omega\omega)\right]\left(\frac{1.5 {\rm
    ~GeV}/c^2}{\rm M}\right)\left(\frac{50  {\rm
      ~MeV}/c^2}{\Gamma_{\eta(1760)}}\right)\frac{45}{39} \\
&\simeq& \frac{98\pm16 \rm MeV}{244^{+35}_{-33} \rm MeV} 
= 0.40_{-0.09}^{+0.08}, \\
\end{eqnarray*}
where $x|H_{PS}(x)|^2$ is taken as 39, which is obtained from 
Fig.1 of Ref.~\cite{close}, and the theoretical uncertainty is not considered.
Since we expect $Br(\eta(1760)\to\omega\omega) < 1.0$, the relationship above  
implies $Br(\eta(1760)\to gg) > 0.28$ at 90\% confidence level.

 From a theoretical viewpoint, the
coupling of a glueball to two photons is expected to be very weak,
so the study of $\eta(1760)$ production in the two photon
process is needed. The $\eta(1760)$ is abundantly produced in the
$J/\psi$ radiative decay, but it is not seen in $J/\psi\to
\gamma \gamma
V(\rho,\phi)$~\cite{radiative1,radiative2,radiative3, radiative4},
which means that the partial width of $\eta(1760) \to \gamma
V(\rho,\phi)$ is very small. $\eta(1760)$ is shown to have large
gluon component, but its mass is much lower than the prediction
from lattice QCD calculation~\cite{lattice}, suggesting that it is a
mixture of the glueball and $q\bar{q}$ meson.  If $\eta(1760)$ is a mixed pseudoscalar
glueball candidate, it should have flavor symmetric decays. Therefore other decay modes of $\eta(1760)$, such as
$\eta(1760) \to \rho \rho $, $K^* K^*$, $\eta \pi \pi$, $K \bar{K}
\pi$, etc. should be studied. Further studies are
needed to understand the nature of the $\eta(1760)$, both experimentally
and theoretically.

\section{ \boldmath Summary}

In summary, $J/\psi\to\gamma \omega\omega$,
 $\omega\to\pi^+\pi^-\pi^0$ is studied, and the $\omega\omega$ invariant
mass distribution peaks at 1.76 GeV/$c^2$. The partial wave
analysis shows that the structure is predominantly pseudoscalar,
with small contributions from $f_0(1710)$, $f_2(1640)$, and
$f_2(1910)$. The mass of the pseudoscalar is $\rm {M} $ = 1744 $\pm$ 10 (stat)
$\pm$ 15 (syst) MeV/$c^2$, the width $\Gamma$ =
$244^{+24}_{-21}$ (stat) $\pm$ 25 (syst) MeV/$c^2$, and the product
branching fraction is Br($J/\psi\to\gamma\eta(1760)$) $\cdot$
Br($\eta(1760)\to \omega\omega$) = (1.98 $\pm$ 0.08 (stat)  $\pm$ 0.32 (syst))
$\times$ 10$^{-3}$. The corresponding product branching fractions with
intermediate resonances $f_0(1710)$, $f_2(1640)$, and $f_2(1910)$
are also determined, but with larger errors.

\section{\boldmath Acknowledgment}
The BES collaboration thanks the staff of BEPC and computing
center for their hard
efforts. This work is supported in part by the National Natural
Science Foundation of China under contracts Nos. 10491300,
10225524, 10225525, 10425523, the Chinese Academy of Sciences under
contract No. KJ 95T-03, the 100 Talents Program of CAS under
Contract Nos. U-11, U-24, U-25, and the Knowledge Innovation
Project of CAS under Contract Nos. U-602, U-34 (IHEP), the
National Natural Science Foundation of China under Contract No.
10225522 (Tsinghua University), and the Department of Energy under
Contract No.DE-FG02-04ER41291 (U Hawaii).


\begin{thebibliography}{**}
\bibitem{pdg} Particle Data Group, S. Eidelman et al., Phys. Lett. 
B $\bf 592$, 1 (2004).
\bibitem{Scharre} D. L. Scharre et al., Phys. Lett. B $\bf 97$ , 
329 (1980).
\bibitem{Behrend} H. J. Behrend et al. (CELLO Collaboration),
Z. Phys. C $\bf 42$, 367 (1989).
\bibitem{radiative1} D. Coffman et al. (MARK-III Collaboration),
Phys. Rev. D $\bf 41$, 1410 (1990).
\bibitem{radiative2} J. E. Augustin et al. (DM2 Collaboration),
Phys. Rev. D $\bf 42$, 10 (1990).
\bibitem{radiative3} C. Edwards, PhD thesis, Cal. Tech. Preprint
CALT-68-1165 (1985).
\bibitem{radiative4} M. Ablikim et al. (BES Collaboration),
Phys. Lett. B $\bf 594$, 47 (2004).
\bibitem{L3} I. Vodopianov (L3 Collaboration),
Acta Phys.  Polon.  B $\bf 31$, 2453 (2000).
\bibitem{lattice}  C. J. Morningstar and M. Peardon,
Phys. Rev. D $\bf 60$, 034509 (1999).
\bibitem{ww-mark} R. M. Baltrusaitis et al. (MARKIII Collaboration),
Phys. Rev. Lett. $\bf 55$, 1723 (1985).
\bibitem{rr-mark} R. M. Baltrusaitis et al. (MARKIII Collaboration),
Phys. Rev. D $\bf 33$, 1222 (1986).
\bibitem{rr-dm} D. Bisello et al. (DM2 Collaboration),
Phys. Rev. D $\bf 39$, 701 (1989).
\bibitem{ww-dm} D. Bisello et al. (DM2 Collaboration),
Phys. Lett. B $\bf 192$, 239 (1987).
\bibitem{bes} J. Z. Bai et al. (BES Collaboration),
Phys. Lett. B $\bf 446$, 356 (1999).
\bibitem{etkin} A. Etkin et al., Phys. Rev. Lett. $\bf 40$, 422 (1978).
\bibitem{li}  P. R. Page, X. Q. Li, Eur. Phys. J. C1, $\bf 579$ (1998).
\bibitem{wu}  N. Wu et al,  Chin. Phys. $\bf 10$, 611 (2001); 
/hep-ph/0011338.
\bibitem{liba} B. A. Li, hep-ph/0510093 (2005).
\bibitem{BESII} J. Z. Bai et al. (BES Collaboration), Nucl. Instr.
Meth. A $\bf 344$, 319 (1994).
\bibitem{simbes} M. Ablikim et al. (BES Collaboration),
Nucl. Instr. Meth. A $\bf 552$, 344 (2005)
\bibitem{chenjc} J. C. Chen et al., Phys. Rev. D $\bf 62$, 034003 (2000).
\bibitem{chi1} N. P. Chang and C. A. Nelson, Phys. Rev. Lett. $\bf 40$, 
1617 (1978).
\bibitem{chi2} T. L. Trueman et al., Phys. Rev. D $\bf 18$, 3423 (1978).
\bibitem{chi3} R. M. Baltrusaitis et al., Phys. Rev. Lett. $\bf 52$, 
2126 (1984).
\bibitem{chi44} D. Bisello et al. (DM2 collaboration), Nucl. Phys. B 
$\bf 350$, 1 (1991).
\bibitem{wuning1} N. Wu and T. N. Ruan, Commun. theor. Phys. (Beijing, China)
$\bf 35$, 547 (2001); \\$\bf 35$, 693 (2001); $\bf 37$, 309 (2002).
\bibitem{guozj} M. Ablikim et al. (BES Collaboration),
Phys. Rev. D $\bf 72$, 092002 (2005).
\bibitem{minuit} CERN Program Library D $\bf 506$.
\bibitem{dongly} M. Ablikim et al. (BES Collaboration),
Phys. Lett. B $\bf 610$, 192 (2005) .
\bibitem{wu1} M. Ablikim et al. (BES Collaboration),
Phys. Lett. B $\bf 598$, 149 (2004).
\bibitem{penghp} M. Ablikim et al. (BES Collaboration), 
                 Phys. Rev. Lett. 96, 162002(2006).
\bibitem{Liba2} B. A. Li, hep-ph/0602072.
\bibitem{ppp}   P.  Bicudo et al., hep-ph/0602172.
\bibitem{zhao}   K. T. Chao, hep-ph/0602190.
\bibitem{bug}   D.V. Bugg, hep-ph/0603018.
\bibitem{hexg1} Xiao-Gang He, Xue-Qian Li et al., Phys. Rev. D 73,
                051502 (2006).
\bibitem{hexg}  Xiao-Gang He, Xue-Qian Li et al., hep-ph/0604141.
\bibitem{close} F. E. Close, G. R. Farrar, Z. Li, 
Phys. Rev. D $\bf 55$, 5749 (1997).
\bibitem{cakir} M. B. Cakir, G. R. Farrar, 
Phys. Rev. D $\bf 50$, 3268 (1994).
\end{thebibliography}
\end{document}